\documentclass[final]{aipproc}
\layoutstyle{6x9}
\begin{document}

\title[Multi-Quark Hadrons]{Multi-Quark Hadrons and S=-2 Hypernuclei.}    
\author{D.~E.~Kahana}{address={Physics Department, Brookhaven National
Laboratory, Upton, NY 11973, USA}}
\author{S.~H.~Kahana}{address={Physics Department, Brookhaven National
Laboratory, Upton, NY 11973, USA}}

\begin{abstract}

\date{\today}  
  
The general  character of  4-quark (mesonic) and  strange 6-quark
(baryonic)  quark systems is  very briefly  reviewed a  la Jaffe,
{\it  i.~e.}~in the  MIT bag,  and so  far still  possibly viable
candidates   are   indicated.    Concentration   is   on   $S=-2$
systems. Traditionally, one employs the $(K^-,K^+)$ reaction on a
relatively  light  target  and  hopes  to  retain  two  units  of
strangeness  on a  single final  state  fragment.  Alternatively,
heavy  ion reactions  can be  used to  produce $\Lambda$-hyperons
copiously and  one seeks to  observe coalescence of two  of these
particles  into the  lightest $S=-2$  nucleus,  the $H$-dibaryon.
The complications  arising from the presence of  a repulsive core
in the baryon-baryon interaction on the production of the $H$ are
discussed.  Also considered is  the possible presence in the data
from the AGS experiment  E906, of slightly heavier $S=-2$ nuclei,
in particular $_{\Lambda\Lambda}^{\,\,\,\,4}$H.

\end{abstract}   

\maketitle  

\section{INTRODUCTION}  

One  very interesting  question which  arises in  our  search for
examples  of quark-gluon matter  is the  apparent absence,  or at
least the  paucity of  examples of elementary  hadrons possessing
more than three quarks.  There  is perhaps one good candidate for
an  exotic  meson consisting  of  two  quarks  combined with  two
anti-quarks:  {\it  viz}  the  $\pi_1$  1410  thought  to  be  an
$I^G(J^{PC}=1^-(1^{-+})$ state \cite{taatables}, but there is not
likely an equally good candidate for  a $gg +ggg + \cdots $ state
or  glueball.    The  perhaps  more   distinguishable  H-dibaryon
\cite{Jaffe}  has  also  not yet  shown  up  on  its own  in  any
experimental search  and it has  proved to be equally  elusive in
theoretical  analysis.  Surely, however,  strange matter  must be
present at  the heart of virtually  all gravitationally collapsed
objects \cite{BCK}.  Nothing new  is offered here with respect to
the mesonic possibilities: but the presence of more than a single
strange quark in  dibaryons and light nuclei is  explored in more
detail.

During  this  presentation,  we  wish  to  cover  two  apparently
disparate subjects: (1) the  production of the H-dibaryon in both
elementary  and   heavy  ion  induced  reactions,   and  (2)  the
generation of very light  to moderately light $S=-2$ hypernuclei.
Both subjects concern $S=-2$  systems and they are quite possibly
tied together by the possible presence within finite systems of a
hybrid    H,   possibly    constructed    both   from    dibaryon
$\Lambda\Lambda$   and  from   $6$-quark   bag  like   ($uuddss$)
components, {\it viz}:

\begin{equation}
\vert \Psi \rangle =  \alpha \vert \Lambda\Lambda \rangle + \beta
\vert q^6 \rangle
\end{equation}
with  $\alpha,  \beta$  being  amplitudes for  the  two-body  and
$6$-quark components of the  hybrid state.  The purely Jaffe-like
H  state  \cite{Jaffe}   corresponds  to  $\beta=1$.   Our  later
coalescence  calculation   for  the   formation  of  an   $H$  is
independent  of  this  parameter.   The procedure  we  follow  to
estimate the effect  of a repulsive core on  entry from a doorway
$\Lambda\Lambda$ state  into the final H is  applicable to either
the pure bag or hybrid cases.

We  begin by  indicating that  the seeming  absence of  the  H in
existing   searches   is   perhaps  attributable   to   repulsive
(soft-core)  forces in  the baryon-baryon  system,  which prevent
penetration to short range  of a $\Lambda\Lambda$ pair during any
mechanism for formation of  the dihyperon structure. However, one
might anticipate  to the contrary,  that within a  finite nucleus
two  $\Lambda$'s could be  held together  for sufficient  time to
permit a short range structure to develop.

Since  the  first  of  these subjects,  H-suppression,  has  been
described  elsewhere  \cite{hardcore}, it  will  only be  briefly
dealt with  here. To our  knowledge all theoretical  estimates of
production   rates  \cite{hlight,KahanaDover},   irrespective  of
mechanism, have overlooked the possibility of a repulsive core in
the baryon-baryon  interaction at  short distances.  As  we show,
under reasonable assumptions the  core can lead to an appreciable
diminution of H yield. We introduce this device in the context of
heavy    ion   collisions    where    a   previous    calculation
\cite{KahanaDover} suggested a  high formation probability, $\sim
0.07$ per  central Au  + Au collision.   The AGS  experiment E896
\cite{Crawford} is  presently analysing some  100 million central
Au + Au events and could, in the light of this previous estimate,
have  provided  a definitive  search  for  the  H.  In  Reference
\cite{hardcore} we presented an estimate of the extent to which a
repulsive core might interfere with this hope.

We   put  forward   first  the   simplest   possible  theoretical
description  of the  multi-quark systems;  then we  consider the,
successful or otherwise,  experimental searches for such objects.
This  is followed  by a  description of  the related  attempts to
produce multi-strange nuclei.

\section{Bag Model Analysis of Multi-Quark States}

Soon after the introduction of the MIT bag model \cite{bag}, used
to consider the normal  hadrons, mesons constructed from a single
quark-antiquark pair  and baryons containing  three quarks, Jaffe
\cite{Jaffe} proposed the insertion into the bag of extra valence
quarks could produce more exotic systems.  For meson states these
additional  components could  be quarks  only $(q  \bar  q)^2$ or
could  be hybrids  of quarks  and glue  $(q \bar  q,g)$. Glueball
states, $(gg + ggg +  \cdots)$, were already on the table.  Jaffe
also  suggested the existence  of the  H-Dibaryon which  could be
described as a $6$-quark bag (uu,dd,ss) \cite{Jaffe}.  Of course,
the easiest  meson candidates to identify would  be the so-called
exotics,  {\it  i.~e.}~those with  quantum  numbers which  cannot
arise  from  ($q \bar  q$)  alone.   Examples  of exotic  quantum
numbers  are states  characterised by  $J^{PC} =  0^{--}, 0^{+-},
1^{-+},2^{+-}$.

The basic  theory covering all these  states in the  bag model is
essentially the same.  The total energy  for N quarks in a bag of
radius R may be written \cite{Jaffe}:

\begin{equation}
E(N) = E_Q + E_V + E_0 + E_G
\end{equation}
\noindent with a quark kinetic energy 
\begin{equation}
E_Q = \frac{1}{R}\sum_{i=1}^N [ k(m_iR)^2 + (m_iR)^2],
\end{equation}

\noindent where $k$ is a wave number, and  
\begin{equation}
E_G  = -\frac{g^2}{2}\sum_{i<j}\sum_{a=1}^8\int  d^3x  \, \left(\bar B_i^a \cdot \bar B_j^a\right) =
    -\frac{\alpha_c}{R}\sum_{i<j}\sum_{a=1}^8 \, M(m_iR,m_jR)\,     
[\bar \sigma_i \cdot \bar \sigma_j] \,\,(\lambda^a_i \lambda^a_j)\,\,\,.  
\end{equation}

Here $M(m_iR,m_jR)$,  is the  color magnetic energy  of one-gluon
exchange from  quark to  quark.  The other  terms $E_V =  BV$ and
$E_0 =-z_0/R $ are the volume bag pressure and the bag zero point
energy respectively.  The bag parameters suggested by analysis of
the normal hadron spectrum  involve a rather large gluon exchange
coupling $\alpha_c=(g^2/4\pi) = 0.55$,  a bag constant $B^{1/4} =
146$  MeV, vacuum  energy $z_0/R  = 1.84$,  and finally  a rather
large mass  for the strange quark,  $m_s = 279$  MeV.  The matrix
element,   $M(m_iR,m_jR)$,   obtained   in   the  bag   for   the
color-magnetic operators is approximated by a diagonal matrix for
strange  quarks;  $M=M((n_s/N)m_sR,(n_s/N)m_sR)$  with $n_s$  the
number of strange quarks.

To  proceed,  one  constructs  anti-symmetric wave  functions  in
color, spin  and flavour, diagonalises the  energy exploiting the
existing $SU(6)=SU_c(3)XSU_S(2)$ symmetry  and then minimises the
whole with  respect to bag  radius.  The general range  found for
the masses of $(q \bar q)^2$ states is from $1460$ MeV ($S=0$) to
$2140$  ($S=4$) MeV.   A reasonably  good candidate  for  such an
exotic is the $\pi_1 (1400)$, referred to above \cite{taatables},
identified in $\pi^- N$  scattering.  In particular Thompson {\it
et  al.}~\cite{taatables}  find a  resonant  final state  $(\eta,
\pi^-)$  component to  which  they  assign a  mass  of $1370  \pm
10^{+50}_{-30}) $  MeV and a  width of $385  \pm 40^{+65}_{-105}$
MeV.   No state  of similar  status has  been isolated  for other
quark hybrids or glueballs.

The $S=-2$ H-dibaryon,  considered more extensively hereafter, is
found by  Jaffe, through reasoning  similar to that above,  to be
some $80$ MeV bound.

\section{H-Dibaryon} 

This highly symmetric object is in principle more likely to exist
than its meson-like compatriots. The H possesses conserved baryon
and strangeness  numbers, {\it  viz} $B=2, S=-2$.   If it  were a
purely hadronic state its wave function would appear as

\begin{equation} 
\Psi_H      = 
\sqrt{\frac{1}{8}}  \vert \Psi_{\Lambda\Lambda}\rangle
+
 \sqrt{\frac{3}{8}} \vert \Psi_{\Sigma\Lambda}\rangle
+
 \sqrt{\frac{4}{8}} \vert \Psi_{\Xi N} \rangle.
\end{equation}

More recent  estimates of  the mass of  the H have  differed from
Jaffe's original estimate and from each other.  Indeeed, there is
no  consensus among  theorists  that the  H  is in  fact a  bound
object.  There  is, however, general agreement on  the results of
the many searches which have been  made for this state: it is not
yet found.

Conventional  hypernuclear   studies  \cite{hlight}  exploit  the
elementary processes $p(K^-,K^+)\Xi^-$, $p(\Xi^-,\Lambda)\Lambda$
The first produces an effective $\Xi^-$ beam, incident on another
nucleus in the target and thus generates a di-lambda.  The latter
pair may or  may not form the putative  H.  The same experimental
approach could  of course  yield doubly strange  hypernuclei.  An
advantageous  target  for  the  $\Xi^-$  beam  is  the  deuteron,
resulting hopefully  in an H  plus a monoenergetic  neutron.  The
latter is relatively easily identified.

Another approach, used in the BNL experiment E896 \cite{Crawford}
exploits the  large numbers of $\Lambda$'s, some  $30$ per event,
generated in  relativistic $Au +  Au$ collisions. Here  again the
results have been so far negative.

\begin{figure}[tbp]
\includegraphics[height=.8\textheight]{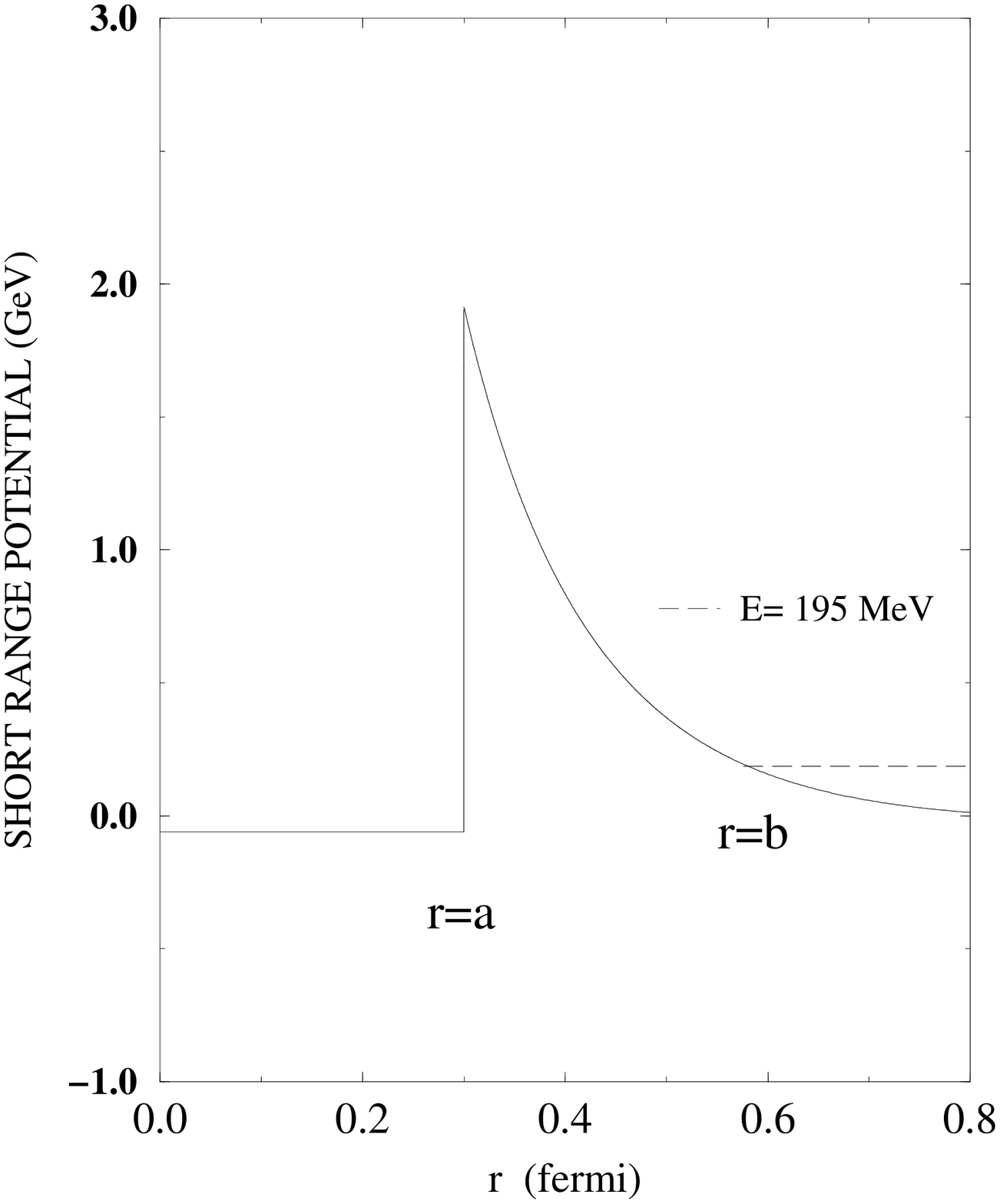}
\caption{The short range  $\Lambda\Lambda$ potential taken from
the $\sigma$  and $\omega$ exchange parts of  the Bonn potential.
At some  separation $a$  the two strange  baryons are  assumed to
dissolve  into  a single  bag,  represented  here  by a  shallow,
attractive potential.  Barrier penetration at relative energy $E$
from  the di-lambda doorway  state is  represented by  the dashed
line.}
\label{fig:one}
\end{figure}

One good reason for the lack of success of these experiments lies
in the nature of the  production of a bound di-lambda state.  For
two  strange baryons  to  coalesce  into a  bag  they must  first
penetrate   the  mutually  repulsive   core  of   the  potential.
Suppression  of the yield  for a  spatially very  extended object
like the  deuteron is minimal.  This  is not so for  the H, which
consists of six  quarks in a bag comparable in size  to that of a
single  baryon, so  that  short range  repulsion,  found in  $NN$
interactions  and  expected to  exist  for strange  baryon-baryon
interactions  as  well,  could   play  a  considerable  role.   H
formation from two $\Lambda$s can  be viewed as proceeding in two
steps:  merging  into a  broad  deuteron-like  state followed  by
barrier  penetration into  the  bag.  The  overall  rate is  then
proportional  to the  product of  the probability  of coalescence
\cite{dekcoal}   with   a   prefactor  giving   the   penetration
probability.    Of   course   there   are   unknowns,   one   the
$\Lambda\Lambda$  separation  $a$ at  which  the  two bags  would
dissolve into  a single  bag, the other  the nature of  the short
range forces after  dissolution. The first, $a$, we  treated as a
parameter;   the  second   we  took   from  the   Bonn  potential
\cite{Bonn}, limiting  our considerations to  the shortest ranged
$\omega$ and $\sigma$ components.  Since the exchange of a $\phi$
meson between $s$ quarks very  nearly matches that of an $\omega$
between  ordinary  quarks,  the  Bonn  interaction  needs  little
readjustment for  strange-strange interaction.  Thus  we take the
short range force, appropriate to the penetration of the core and
depicted in Fig. \ref{fig:one}, to be of the form

\begin{equation}
V(r)=V_{\omega}(r)+V_{\sigma}(r),
\end{equation}
where
\begin{equation}
V_i(r)= \frac{g_i}{r} \exp (-m_ir). 
\end{equation}

The strong short range  $\sigma$ attraction reduces the effect of
the  hard core, while  the longer  range parts  of the  force are
assumed to  play a negligible  role. The two baryons  approach to
some  outer radius  $a$, in  fact a  turning point,  before being
faced with  the strong repulsion.  The  calculation is especially
sensitive to  this separation $a$. Although our  final results on
barrier penetration are  consequently somewhat uncertain, it will
become apparent  that the  one thing one  cannot do is  to ignore
them.

Barrier  penetration  in an  effective  two  body  model can  the
quantified  in  the  transmission coefficient  \cite{Messiah}  at
relative energy $E$:

\begin{equation}
T(E)= 4 \exp (-2\tau),
\end{equation}
where
\begin{equation}
\tau=  \int_a^b\,dr\,\sqrt{2m(V(r)-E}.
\end{equation}

\begin{figure}[tbp]
\includegraphics[height=.8\textheight]{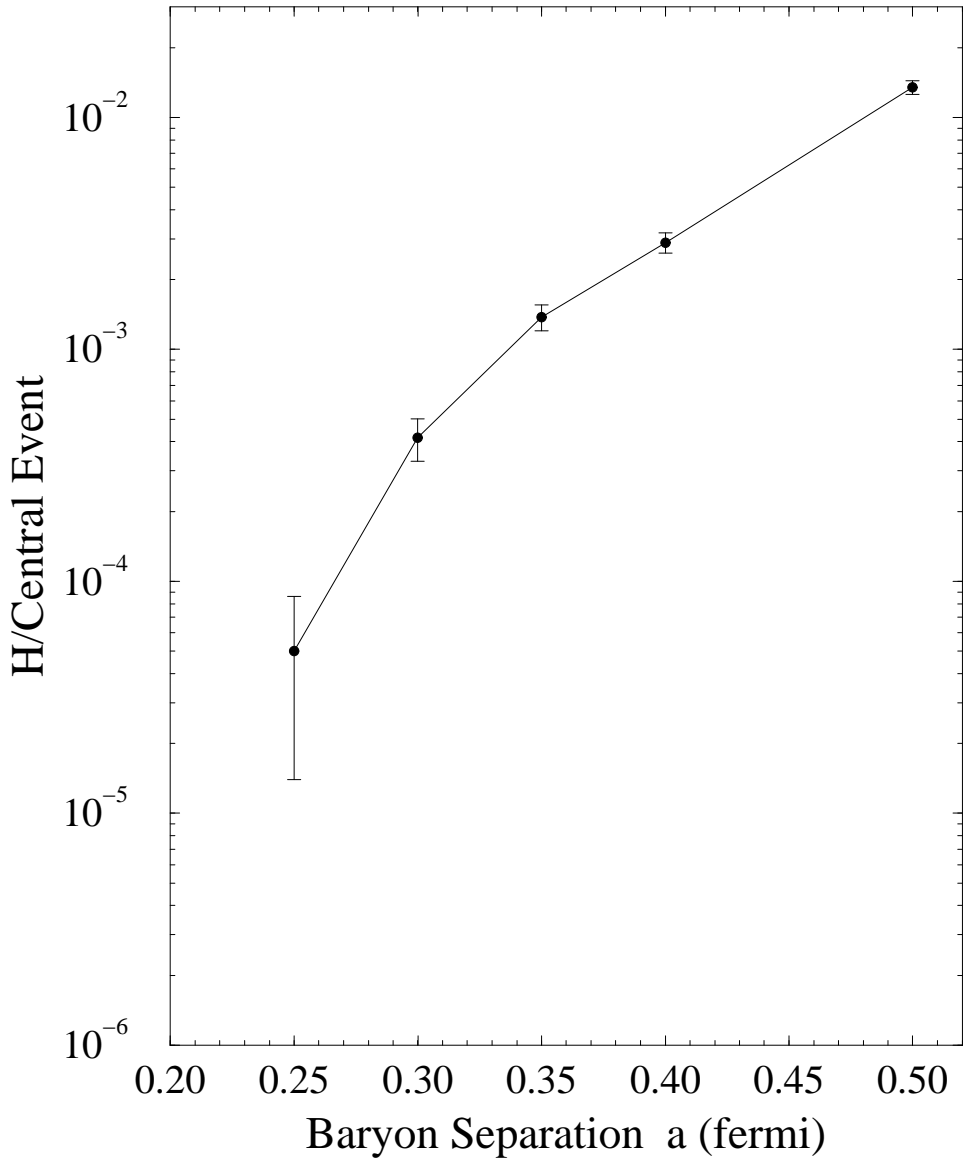}
\caption{Absolute H  Production per Central  Au+Au collision at
$10.6$  GeV.  The  precise separation  of $\Lambda$  bag centers,
$a$,  at which  a single  six quark  bag forms  is of  course not
known, but a reasonable value  is likely less than $0.3$ fm. Even
for  complete  baryon   overlap,  the  average  distance  between
constituent   quarks    is   greater   than    the   conventional
nucleon-nucleon hard core radius of $0.4$ fm.}
\label{fig:two}
\end{figure}

The chief results, as  they relate to coalescence into di-baryons
in  Au +  Au collisions,  are presented  in  Fig.  \ref{fig:two},
indicating  the variation  of  H-dibaryon yield  with $a$.   This
latter parameter must not be thought of as an effective hard core
radius  for  the  $\Lambda\Lambda$ interaction.   The  underlying
quark-quark forces may be  viewed as possessing a repulsive short
range   component  due   to   the  exchange   of  vector   mesons
\cite{kahanaripka}.   Even with  complete overlap  of  the parent
baryons the  average interquark separation for  a uniform spatial
distribution  is comparable  to the  parent radius  $R  \sim 0.8$
fermis,  {\it i.~e.}~considerably  greater  than any  conceivable
fixed hard core.  We  have concentrated on baryon centers between
$0.25$ and $0.40$ fermis apart as a reasonable range.

At the largest  $a$ suppression from the repulsive  forces is not
inconsiderable,  but for the  smallest $a$-values  observation of
the H, should it exist, becomes difficult.  Early analysis of the
actual  experimental setup  using  the heavy  ion simulation  ARC
\cite{arc,Judd}, suggested  a neutral background  comparable with
the initial estimate of  $0.07$ $H$'s per central collision.  For
baryon separations of  $0.25$ to $0.35$ fermis one  would have to
achieve a tracking sensitivity of $10^{-4}$ to $10^{-2}$ relative
to background.  Even in the worst case scenario one is still left
with perhaps a few thousand sample dibaryons, from the very large
number of central  Au + Au events, but they  are immersed in what
may prove to be a daunting background.

We conclude  that short  range repulsion between  strange baryons
can  profoundly  hinder   coalescence  into  objects  whose  very
existence   depends  on  the   presence  of   important  bag-like
structure.  This lesson applies  even more to the many H-searches
initiated  in $(K^-,K^+)$  reactions  \cite{hlight}, since  these
generally   involve    even   lower   relative    energies   $E$.
Unfortunately   the  very  same   repulsive  forces   which  made
coalescence into a bound state  difficult, may also, at the quark
level, destroy the existence of any such state.

\section{Doubly Strange Hypernuclei}

There is  perhaps one way to circumvent  this frustrating barrier
to the discovery of the  lightest of all possible strangelets. In
the event a pair of $\Lambda$s is attached, through a $(K^-,K^+)$
reaction,  to  a  light  nucleus,  a  hybrid  H  may  form,  {\it
i.~e.}~the  H described  above containing  both dibaryon  and six
quark  bag   components.   In   a  light  nucleus,   for  example
$_{\Lambda\Lambda}^{\,\,\,\,4}$H, the extra nucleons in this four
particle  system keep  the  captured hyperons  together for  some
$100$ picoseconds,  far more than enough time  for penetration of
the rather  modest $1-2$ GeV  barrier between them.  In  any case
the search  for $S=-2$ hypernuclei,  the only strangelets  we are
certain exist, is highly interesting in its own right. An ongoing
AGS experiment, E906 \cite{E906prl},  focuses on these nuclei and
as  we indicate  may  have already  provided  evidence for  their
existence \cite{E906prl}.

In  E906  \cite{E906prl} the  ($K^-,K^+$)  reaction  on a  $^9$Be
target is  employed to  produce a tagged  beam of  $\Xi^-$s.  The
$\Xi^-$  in  turn  may  convert  into a  pair  of  $\Lambda$s  by
interaction with a proton, either  in the nucleus in which it was
produced  or  by   subsequent  interaction  with  another  $^9$Be
nucleus.  It is in an emulsion experiment at KEK \cite{Aoki} that
one such  hypernuclei was found with perhaps  other examples from
Prowse \cite{Prowse} and Danysz et al \cite{Danysz}.  Indeed, all
three of these candidates for double hypernuclei were interpreted
as   possessing  a   $\Lambda\Lambda$   pairing  energy   $\Delta
B_{\Lambda\Lambda} \sim  4.5$ MeV  \cite{Millener} . Such  a high
value is unexpected  from existing, albeit theoretical, knowledge
of  hyperon-hyperon  forces.  It  was  then  possible to  surmise
interesting  activity  in the  $\Lambda\Lambda$  system at  short
range separation.

Given the  $K^-$ beam  energy of $1.8$  GeV, the  tagged $\Xi^-$s
initially  possess considerable kinetic  energy, $\sim  140$ MeV,
and are more likely than not  to escape the nucleus in which they
are  produced.  A  guiding principle  we employ  in qualitatively
understanding the broad aspects of the data, is that final states
containing the very stable  $^4$He are favoured.  This picture is
also  mindful of  an oft  used cluster  model for  $^9$Be  as two
$\alpha$-particles  joined together  by a  weakly  bound neutron.
The $^9$Be target is useful  in slowing down the $\Xi^-$, but the
observed reactions are for the most part initiated essentially on
an $\alpha$.

Further, in  the present  experiment which observes  $S=-2$ final
states by their decay into two momentum-correlated $\pi^-$'s, the
heaviest   $A=9,8,7,\cdots   $    systems   will   decay   weakly
predominantly  through non-mesonic  channels.  Thus  the lightest
systems are  probably doubly favoured  in our data,  through both
their production and their decay dynamics.

\section{Analysis of the E906 Data}

\begin{figure}[tbp]
\includegraphics[height=.5\textheight]{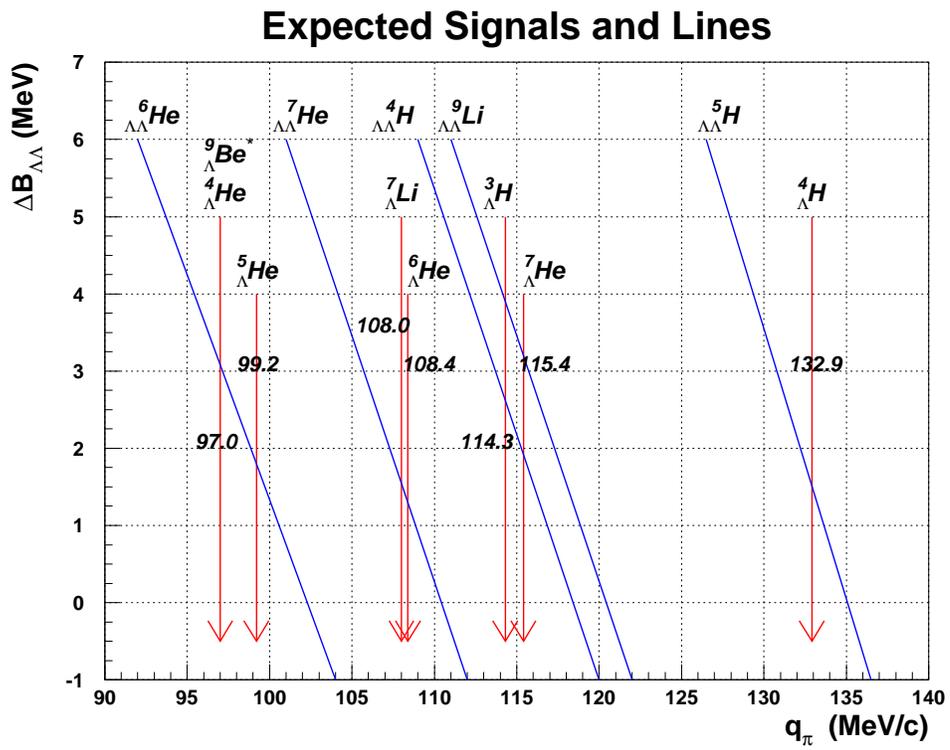}
\caption{Expected $S = -1$  lines together with possible doubly
strange hypernuclear energies, the  latter given as a function of
the pairing energy $\Delta B_{\Lambda\Lambda}$.}
\label{fig:4_906}
\end{figure}

The   measurements   which    contain   evidence   for   possible
$\Lambda\Lambda$    hypernuclei   are    highlighted    in   Fig.
\ref{fig:3_906}. Displayed  are the two-dimensional,  Dalitz plot
for the total sample of correlated two $\pi^-$ decays recorded in
the CDS detector \cite{E906prl}, and the projections of this data
on  the $P_L$  and  $P_H$ axes.  These  denote the  low and  high
momenta for  the $\pi_L$ and  $\pi_H$ pair detected in  an event.
The circled  feature in  the 2-D plot  is of  principle interest,
giving  rise to the  prominent peaks  near $P_H  = 114$  MeV/c in
insert I and a correlated peak at $P_H \sim 104$ MeV/c in II. The
experiment  is searching  for pairs  of pion  momenta unexplained
from  previous  knowledge  of  hypernucleon lines.  The  expected
spectrum close  to the  region of interest  in E906 are  shown in
Fig.  \ref{fig:4_906},  with single hypernuclear  lines indicated
vertically and doubly strange candidates diagonally. The slope in
the  latter denotes  the dependence  of  the $S=-2$  energy as  a
function    of    the    important   pairing    energy    $\Delta
B_{\Lambda\Lambda}$.

The higher momentum structure in I near $137$ MeV/c is understood
as the decay in flight of  $\Xi^-$ hyperons, but the width of the
lower peak  is too broad  to be due  to a single  component line.
The very prominent peak in II is completely unexpected.  The high
momentum  $\pi^-$ peak  in I  arises in  part from  the  decay of
$_{\Lambda}^{\,\,\,\,3}$H yielding a meson line at $114.3$ MeV/c,
but considering  the experimental  resolution of $2.5$  MeV, this
$\pi_H$ prominence at $114$ MeV must contain more structure.

The projection in II is constructed from a reverse cut $106 MeV/c
\le  k^{\pi^-}  \le  120$  MeV/c,  {\it  i.~e.}~under  the  first
compound  peak in  II, and  thus should  reveal the  low momentum
$\pi^-_L$ correlated  with the $114$  MeV/c peak in I.   The most
striking feature in II is the narrow prominence near $104$ GeV/c,
which  cannot for example  be accounted  for by  single $\Lambda$
decay  in  a $_{\Lambda}^{\,\,\,\,3}{\rm  H}$  + $\Lambda$  final
state, or for that matter by  any other known line. Adding even a
small  kinetic  energy to  the  $\Xi^-$  initiating the  reaction
leading  to  the final  $S=-2$  state  would  broaden the  single
$\Lambda$ decay  well beyond that  measured for the  the dominant
peak in  this reverse  cut.  The correlation  of this peak  in II
with the excessively broad dominant peak in Fig.  \ref{fig:3_906}
I is strong evidence for the presence of a light double-$\Lambda$
species in the data.

Fig.   \ref{fig:4_906},  \cite{Yamamoto},  indicates where  known
single $\Lambda$ hypernuclear lines are expected as well as where
$S=-2$ lines are anticipated as  a function of the pairing energy
$\Delta  B_{\Lambda\Lambda}$. Reiterating,  the  most interesting
feature in  the present data, centered at  $112-114$ MeV/c, could
only       result        from       the       production       of
$_{\Lambda\Lambda}^{\,\,\,\,4}$H and/or $_{\Lambda}^3$H.

$_{\Lambda\Lambda}^{\,\,\,\,4}$H has two generic modes of decay:
 
\begin{equation}
_{\Lambda\Lambda}^{\,\,\,\,4}{\rm      H}      \rightarrow     \,
_{\Lambda}^4{\rm He} + \pi^-_2,
\end{equation}
or
\begin{equation}
_{\Lambda\Lambda}^{\,\,\,\,4}{\rm      H}      \rightarrow     \,
_{\Lambda}^3{\rm H} + p + \pi^-_1.
\end{equation}

The  two body  decay in  the first  mode yields  the  high energy
$\pi^-_H$ from which the $_{\Lambda\Lambda}^{\,\,\,\,4}$H $\Delta
B_{\Lambda\Lambda}$  can  in   principle  be  estimated,  and  is
followed by a three body decay producing the correlated $\pi_L$.

The very narrowness of the $\pi^-_L$ peak in II suggests that the
decay mode in Equation 7 is not strictly three body in character.
We     offer     as     a     candidate    a     resonance     in
$_{\Lambda}^{\,\,\,\,4}$He, arrived at  by decay from the clearly
spatially  extended  $_{\Lambda\Lambda}^{\,\,\,\,4}{\rm  H}$  and
thus not favoured in production from $K^- + ^{\,\,\,\,4}{\rm H}$.
The first, lower, $\pi^-_1$ momentum from the initial decay would
then be  sensitive to the sum $E_R  + \Delta B_{\Lambda\Lambda}$.
Preliminary  theoretical calculations \cite{K2DJM}  indicate that
indeed  a narrow  $p$-wave resonance  can be  placed near  $E_R =
0.5-1.5$ MeV in the  p + $_{\Lambda}^{\,\,\,\,3}$H system , using
a  potential  consistent with  the  rather  extended geometry  of
$_{\Lambda}^{\,\,\,\,3}$H  and constrained  by  the known  proton
binding in the ground state of $_{\Lambda}^{\,\,\,\,4}$He.

\section{Comments and Conclusions}

There is  little to  add to the  above discussion  of multi-quark
elementary  systems.    The  search   for  the  H   is  certainly
complicated by  our analysis of  the mechanism of  its formation.
The  discovery and  study  of $S=-2$  hypernuclei will,  however,
continue either at the AGS  or at the anticipated Japanese Hadron
Facility.

\begin{figure}[tbp]
\includegraphics[height=.5\textheight]{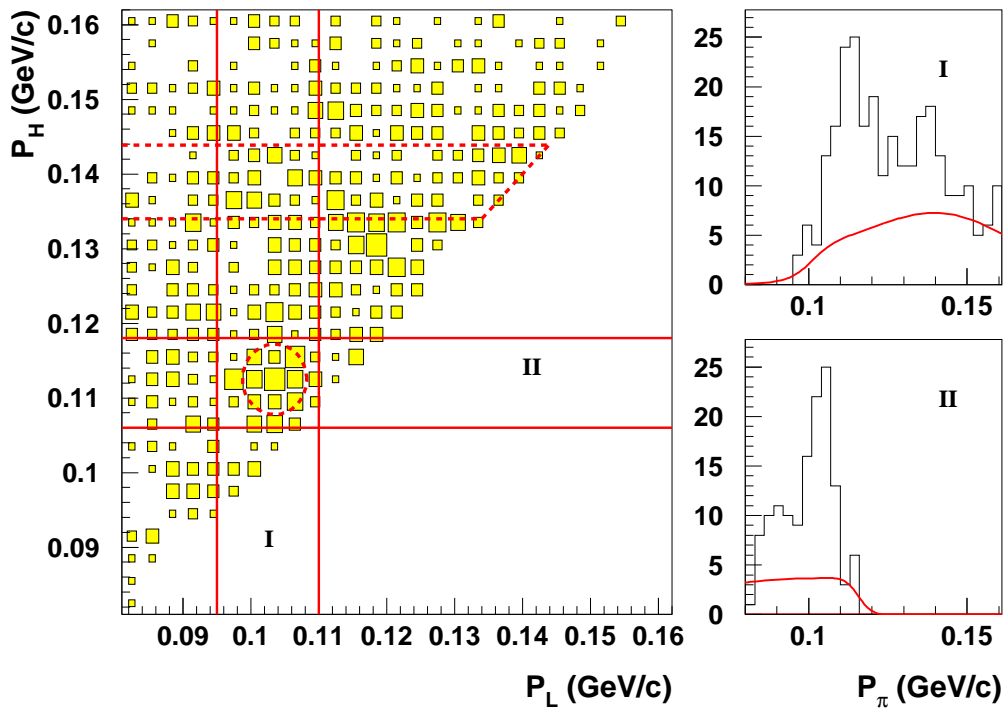}
\caption{E906  Two Pion  Data.  The Dalitz  plot together  with
projections onto high (inset I)  and low (inset II) momentum axes
describe the  approximately $1000$ recorded,  correlated, $\pi^-$
events. The  interesting region  is circled in  the 2-D  plot.  I
contains   a   broad    $\pi_H$   peak   with   components   from
$_{\Lambda}^{\,\,\,\,3}$H    and    the    two-body   decay    of
$_{\Lambda\Lambda}^{\,\,\,\,4}$H.     Inset   II    contains   an
unexpected  narrow $\pi_L$  peak interpreted  as arising  from an
alternative decay of $_{\Lambda\Lambda}^{\,\,\,\,4}$H involving a
resonance in $_{\Lambda}^{\,\,\,\,3}He$.}
\label{fig:3_906}
\end{figure}

For E906 there were $1040$  events in the total two prong sample,
some   $70$   above   background   in  the   first   feature   in
Fig. \ref{fig:3_906}  I.  The  reverse cut in  II can be  used to
isolate        the       three        body        decays       of
$_{\Lambda\Lambda}^{\,\,\,\,4}{\rm H}$ and from the simulation an
estimate of the two body,  resonant, decays can also be made.  We
conclude      then     a      lower     limit      of     $40+20$
$_{\Lambda\Lambda}^{\,\,\,\,4}$H have been  produced in these two
modes  respectively.    Unfortunately,  the  important  di-lambda
pairing energy is not  well determined in the present experiment.
A  simulation   of  Fig.   \ref{fig:3_906}   I  suggests  $\Delta
B_{\Lambda\Lambda}   \sim   1-2$   MeV   while   that   in   Fig.
\ref{fig:3_906}  II  involving  the  combination  with  $E_R$  is
perhaps described by a value closer to $0.5-1.0$ MeV. More recent
emulsion  experiments \cite{newkek} have  definitively identified
the species $_{\Lambda\Lambda}^{\,\,\,\,6}$He  and agree with the
lower  values of  $\Delta B_{\Lambda\Lambda}$,  finding something
less than  $1$ MeV. The motivation  for a hybrid  H living within
some light nucleus then recedes.

Clearly, better  statistics and an improvement  in resolution are
required  to   determine  the  $\Lambda\Lambda$   pairing  energy
definitively,  and  more  importantly  to  establish  the  actual
presence of the large numbers of light doubly strange hypernuclei
suggested by E906. The principal issues to be settled, presumably
in a follow on experiment, are  the existence or not of more than
one contribution to  the broad lower peak in  the higher momentum
$\pi^-_L$ spectrum  and the  very exciting possibility  that more
than   one   species  of   double   $\Lambda$   resides  in   the
data. Significantly higher counting  rate should allow cuts to be
placed on the $\Xi^-$ kinetic  energy and determine the nature of
the  reaction  mechanism   producing  individual  doubly  strange
species.

\section*{ACKNOWLEDGEMENTS}
This manuscript has been authored  under US DOE contracts 
and DE-AC02-98CH10886.

\clearpage

\clearpage

\clearpage

\clearpage

\end{document}